\documentstyle[11pt]{article}
\input epsf
\newlength{\awidth}
\newlength{\aheight}
\newlength{\uswidth}
\newlength{\usheight}
\newlength{\spacing}
\newlength{\topoff}
\newlength{\margoff}
\newlength{\margin}
\newlength{\hmargin}
\newcounter{fignr}
\newenvironment{fig}[1]{\refstepcounter{fignr}\label{#1}\begin{center}}{
    \end{center}}
\newcommand{\figcap}[2]{\parbox{#1}{{\footnotesize  Fig. \thefignr. #2}}}

\begin{document}
\begin{titlepage}
\begin{flushleft}
Stockholm\\
USITP 96-4\\
April 1996\\
\end{flushleft}
\vspace{1cm}
\begin{center}
{\Large MAKING ANTI-DE SITTER BLACK HOLES}\\
\vspace{15mm}
{\large Stefan \AA minneborg}\footnote{Email address: stefan@
vanosf.physto.se}\\
\vspace{5mm}
{\large Ingemar Bengtsson}\footnote{Email address: ingemar@vana.
physto.se}\\
\vspace{5mm}
{\large S\"{o}ren Holst}\footnote{Email address: holst@vanosf.
physto.se}\\
\vspace{5mm}
{\large Peter Peld\'{a}n}\footnote{Email address: peldan@vanosf.
physto.se}\\
\vspace*{5mm}
{\sl Fysikum\\
Stockholm University\\
Box 6730, S-113 85 Stockholm, Sweden}\\
\vspace{15mm}
{\bf Abstract}\\
\end{center}
\
\noindent It is known from the work of Ba\~{n}ados et al. that a 
space-time with event horizons (much like the Schwarzschild black hole)
 can be obtained from 2+1 dimensional anti-de Sitter space through a 
suitable identification of points. We point out that this can be done
 in 3+1 dimensions as well. In this way we obtain black holes with 
event horizons that are tori or Riemann surfaces of genus higher than 
one. They can have either one or two asymptotic regions. Locally, the 
space-time is isometric to anti-de Sitter space.  \\

\end{titlepage}

\noindent It came as a surprise when Ba\~{n}ados et al. produced a
 ``black hole'' solution of Einstein's equations, with a negative
 cosmological constant, in 2+1 dimensions \cite{Banados}. This was
 unexpected because all such solutions are locally isometric to 
anti-de Sitter space, which has constant curvature. Indeed the solution
 can be obtained by a suitable identification of points in anti-de 
Sitter space. The original papers spawned a rather large literature 
(reviewed recently by Carlip and by Mann \cite{Carlip}), but it appears
 to have gone unnoticed that the construction can be generalized to 
higher dimensions, in particular to four dimensions. Our purpose here
 is to remedy this deficiency. It will become evident as we proceed 
that the essential ingredient which makes the construction possible is
 the peculiar asymptotic structure of anti-de Sitter space, which has
 a timelike boundary at spatial infinity. The dimension of space-time
 is not essential.

We wish to acknowledge the work of Brill and Steif \cite{Brill}, who
 stressed that it is helpful to look at the BHTZ construction from an
 initial data point of view.

Before we begin our construction we give a thumbnail sketch of anti-de
 Sitter space. It is defined as the surface

\begin{equation} X^2 + Y^2 + Z^2 - U^2 - V^2 = - 1 \end{equation}

\noindent embedded in a five dimensional flat space with the metric

\begin{equation} ds^2 = dX^2 + dY^2 + dZ^2 - dU^2 - dV^2 \ . 
\end{equation}

\noindent This is a solution of Einstein's equations with the 
cosmological constant ${\Lambda} = - 3$. Its intrinsic curvature 
is constant and negative. We find it
 helpful to think sometimes in terms of the coordinates in the 
embedding space, and sometimes in terms of the intrinsic coordinates
 $(t, {\rho}, {\theta}, {\phi})$, where \cite{Holst}

\begin{equation}
  \label{korvkoo}
\begin{array}{lcr}
\begin{array}{lll}
 X & = &  \frac{2\rho}{1 - \rho^2}\sin{\theta}\cos{\phi} \\
 Y & = &  \frac{\normalsize 2\rho}{\normalsize 1 - \rho^2} \sin{\theta}\sin{\phi} \\
 Z & = &  \frac{2{\rho}}{1 - {\rho}^2}\cos{\theta} \\
 U & = &  \frac{1 + {\rho}^2}{1 - {\rho}^2}\cos{t} \\
 V & = &  \frac{1 + {\rho}^2}{1 - {\rho}^2}\sin{t}
\end{array} &
\mbox{\hspace{1cm}}&
\begin{array}{l}
 0 \leq {\rho} < 1 \\
 0 \leq {\phi} < 2{\pi} \\
 0 \leq {\theta} \leq {\pi} \\
 0 \leq t < 2{\pi}\ . 
\end{array}
\end{array}
\end{equation}

\noindent Most of our reasoning will employ the embedding coordinates,
 but we will use the intrinsic coordinates for drawing the pictures.
 The intrinsic coordinates cover all of space-time, and in terms of them
 the intrinsic metric is

\begin{equation} ds^2 = - \left(\frac{1 + {\rho}^2}
{1 - {\rho}^2}\right)^2 dt^2 + dl^2 \ , \end{equation}

\noindent where 

\begin{equation} dl^2 = \frac{4}{(1 - {\rho}^2)^2}
(d{\rho}^2 + {\rho}^2d{\theta}^2 + {\rho}^2\sin^2{\theta}d{\phi}^2) \
 . \end{equation}

\noindent The metric $dl^2$ is the metric on hyperbolic three-space 
represented as the interior of a unit ball. We refer to this ball as
 the Poincar\'{e} ball, since it is
 the generalization to three dimensions of the Poincar\'{e} disk as a
 model for Lobachevskian geometry (which was reviewed for physicists 
by Balasz and Voros \cite{Balasz}). Hyperbolic three-space  can be 
defined as one sheet of the hyperboloid

\begin{equation} X^2 + Y^2 + Z^2 - U^2 = - 1 \end{equation}

\noindent embedded in flat Minkowski space. In our coordinate system
 anti-de Sitter space has been foliated with Poincar\'{e} balls having
 zero extrinsic curvature.

Some elementary facts about hyperbolic three-space will be used below.
 Its geodesics are segments of circles orthogonal to the boundary of 
the Poincar\'{e} ball. Its isometries are elements of $SO(3,1)$ which 
we call 
rotations and boosts, using a terminology familiar from the study of
 the Lorentz group. A boost can be characterized as an isometry that
 has two fixed points, both situated on the boundary of the ball. Some
 elementary facts about anti-de Sitter space will also be needed, in
 particular the fact that a light ray that passes the spatial origin
 at time $t = 0$ will strike spatial infinity at $t = {\pi}/2$. 
Conversely, the future domain of dependence of the hypersurface $t = 0$
 ends at $t = {\pi}/2$, because of information leaking in from infinity. 

We now turn to a description of the black hole found by Ba\~{n}ados et
 al. \cite{Banados}. They work in 2+1 dimensional anti-de Sitter 
space, which can of course be obtained as the intersection of the 
hypersurface $Z = 0$ with the four dimensional space-time given 
above. To obtain their black hole (more precisely what they call 
their spinless black hole, and what we will call the BHTZ space-time)
 one will have to identify points that can be connected with each 
other by an isometry generated by the Killing vector

\begin{equation}
  \label{killiso}
  J_{XU} = X\partial_U + U\partial_X \ . \end{equation}

\noindent The Killing vector field is time-like in a part of space-time. 
The ``identification surfaces'' are chosen in such a way that there are 
no closed time-like curves in the solution, which means that they should 
lie entirely within the region where the Killing vector field is
 space-like. This we will call the ``allowed region'' --- the covering 
manifold of the BHTZ space-time. It is given by

\begin{equation} U^2 > X^2 \ . \end{equation}

\ \\
\noindent\parbox{\textwidth}{\noindent\begin{fig}{fig1}
  \mbox{\setlength{\epsfxsize}{.80\textwidth} \epsfbox{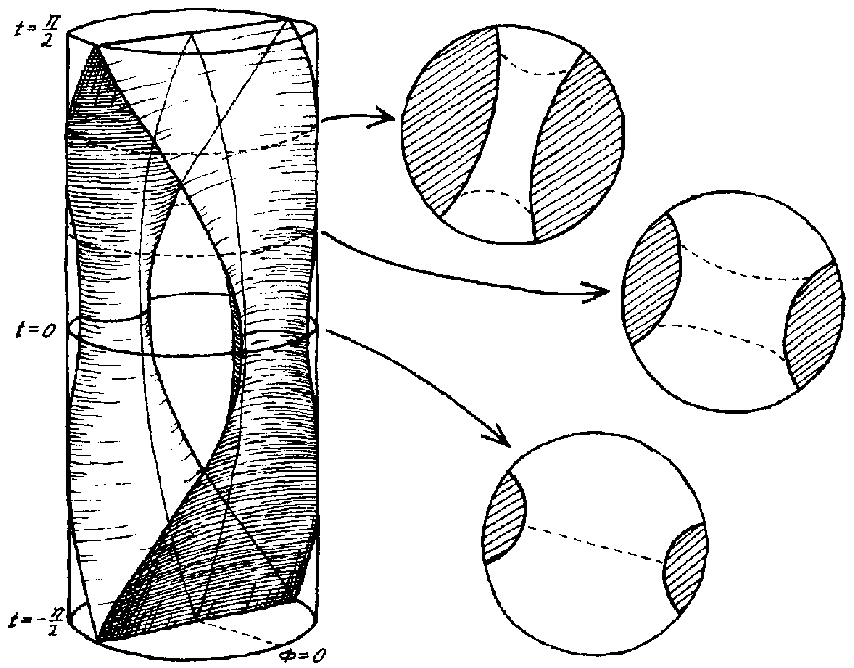} }\\
\ \\
  \figcap{.9\textwidth}{The 2+1 dimensional (spinless) BHTZ solution in coordinates
    ($t$, $\rho$, $\phi$) defined in
    (\ref{korvkoo}) (where $\theta = {\pi}/2$
     since $Z
    = 0$). All points inside the cylinder belong to anti-de Sitter
    space, its surface ($\rho=1$) representing spatial infinity.
    The BHTZ spacetime lies between the two surfaces inside the
    cylinder which are identified under an isometry generated by
    (\ref{killiso}). Since this isometry has fixpoints at $\phi = \pi/2$, $t
    = \pm \pi/2$ the surfaces merge and we have singularities there.
    The future singularity is hidden by an event horizon which ``splits
    up'' at $t = 0$. This horizon is indicated by the dashed lines in
    the constant time slices to the right.}
\end{fig}
}\\

\noindent In figure~\ref{fig1} we have depicted a pair
 of suitable surfaces. They can be obtained by moving the ``vertical''
 hypersurface $X = 0$ backwards or forwards along the Killing vector
 field, and are given by the equation

\begin{equation} \frac{X}{U} = \tanh{u} \label{13} \end{equation}

\noindent for suitable values of the constant $u$. Identifying 
corresponding points on these surfaces gives us the BHTZ space-time.
 The region bounded by the identification surfaces is a regular 
solution of Einstein's equations which is locally isometric to anti-de
 Sitter space. However, when the surfaces merge (at $t = \pm {\pi}/2$)
 the quotient space becomes singular and the BHTZ space-time ends 
there. The singularities are of the ``Misner type'' \cite{Misner}
 --- they are clearly not curvature singularities. There are two 
asymptotic regions in the directions of positive and negative $Y$,
 and the space-time topology is ${\bf R}^2 \otimes {\bf S}^1$.

To see why this is a black hole, consider a light ray that starts out
 from the origin at time $t = 0$. As we observed above, this light 
ray will strike spatial infinity at $t = {\pi}/2$. If we look into 
the BHTZ space-time from the asymptotic region lying in the positive
 $Y$ direction nothing that passes the $t = 0$ hypersurface with a 
negative $Y$ value can be seen --- we have an event horizon and 
therefore a black hole. The location of the event horizon at three 
different times are shown by the dashed lines on the spatial slices
 depicted in figure~\ref{fig1}. Evidently, the Penrose diagram is that 
drawn in figure~\ref{fig2}. Note that this is the same Penrose diagram
 as that of the Schwarzschild-anti-de Sitter solution. The Penrose 
diagram of anti-de Sitter space is also shown, for comparison.

\ \\
\noindent\parbox{\textwidth}{\noindent\begin{fig}{fig2}
  \mbox{\setlength{\epsfxsize}{.5\textwidth} \epsfbox{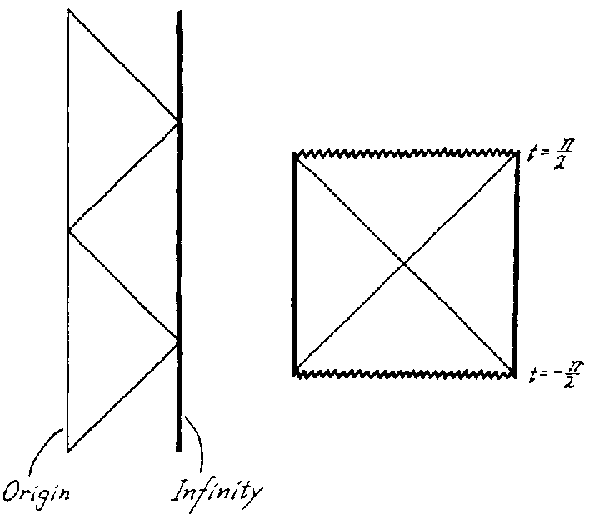} }\\
\ \\
  \figcap{.9\textwidth}{The Penrose diagrams of anti-de Sitter space and the BHTZ
    solution respectively. Note that the latter has two disconnected
    infinities.}
\end{fig}
}\\

\noindent We are now ready to study the situation in four dimensions. In our first
 construction we simply rotate the BHTZ space-time around the $X$-axis.
The identification surfaces are still given by eq. (\ref{13}). Again 
the two surfaces merge at $t = \pm {\pi}/2$, so our space-time begins 
and ends in singularities at these times. To visualize the resulting 
space-time, figure~\ref{fig3} may be helpful. It shows the location of the 
identification surfaces in the Poincar\'{e} balls at three different 
values of $t$ (the first picture is taken at $t = 0$). The main new 
feature compared to 2+1 dimensions is that spatial infinity is 
connected --- there is only one asymptotic region in 3+1 dimensions.

\ \\
\noindent\parbox{\textwidth}{\noindent\begin{fig}{fig3}
  \mbox{\setlength{\epsfxsize}{.9\textwidth} \epsfbox{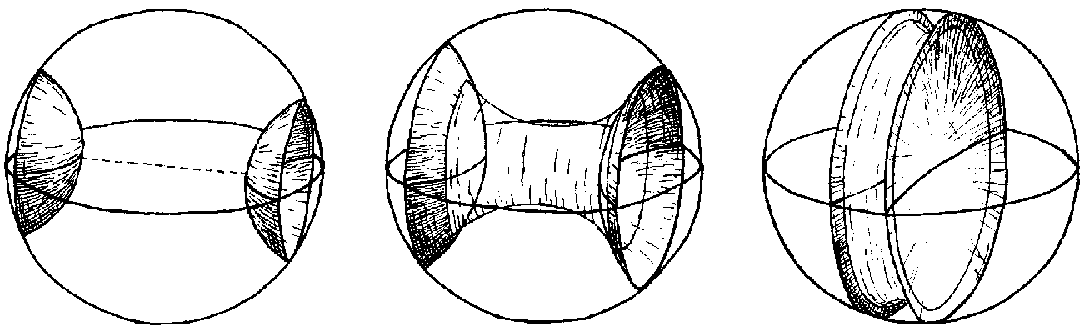} }\\
\ \\
  \figcap{.9\textwidth}{The 3+1 dimensional generalization of the BHTZ solution in
    the coordinates ($\rho$, $\theta$, $\phi$) defined in
    (\ref{korvkoo}) at
    three different times. In these coordinates, at each time, space is
    a Poincar\'{e} ball and the identification surfaces segments of
    spheres, meeting the boundary at right angles. As before an event
    horizon grows up at $t = 0$. In fact, these pictures are obtained
    simply by rotating the constant time slices in figure~\ref{fig1}
    around $\phi = 0$.}
\end{fig}}
\ \\

\noindent\parbox{.5\textwidth}{\noindent Now watch the dashed line ---
actually it is a circle --- that connects
 the identification surfaces at $t = 0$. Photons emitted from this line
 will reach spatial infinity at $t = {\pi}/2$, which is the time when
 space-time ends in a singularity. It is clear that a toroidal event 
horizon will grow up from this line, as shown in the two subsequent 
Poincar\'{e} balls in figure~\ref{fig3}. This is quite similar to the 2+1 
dimensional solution, but due to the fact that there is now only one
 asymptotic region there is also an important difference. The Penrose
 diagram is given in figure~\ref{fig4}. Unlike its 2+1 dimensional
 counterpart this is not an eternal black hole.
}
\parbox{.5\textwidth}{\begin{fig}{fig4}
\ \\
  \mbox{\setlength{\epsfxsize}{.25\textwidth} \epsfbox{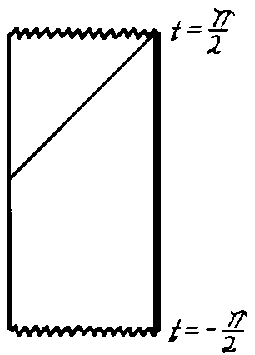} }\\
\ \\
  \figcap{.45\textwidth}{The Penrose diagram of the 3+1 dimensional BHTZ solution.
In contrast to the 2+1 dimensional case, depicted in
    figure~\ref{fig2}, we now have only one infinity.
}
\end{fig}
}\\

Our next example is less trivial. We will identify points that 
can be connected by a discrete subgroup ${\Gamma}$ of
 the $SO(2,1)$ group of isometries generated by the Killing 
vectors $J_{XU}, J_{YU}$ and $J_{XY}$. The ``identification surfaces''
 must lie in the region where the three Killing vector fields are 
spacelike, that is to say that the allowed region is defined by

\begin{equation} U^2 > X^2 + Y^2 \ . \label{14} \end{equation} 

\noindent It is clear that the hypersurface defined by eq. (\ref{13}),
 and hypersurfaces obtained from it by performing rotations generated 
by the Killing vector $J_{XY}$, are suitable choices. The solution will 
then necessarily be singular at $t = \pm {\pi}/2$, because at those 
times every identification surface will form a plane containing the 
$Z$-axis and going straight 
through the middle of the Poincar\'{e} ball.

It is necessary to exercise some care to ensure that these are the 
only singularities that arise. Consider first a simpler case, that 
of a hyperbolic plane ${\bf H}^2$ defined by

\begin{equation} \hat{X}^2 + \hat{Y}^2 - \hat{U}^2 = - 1 \ . 
\label{16} \end{equation}

\noindent It is well known \cite{Balasz} that one can select a discrete
 subgroup ${\Gamma}$ of boosts in $SO(2,1)$ such that the quotient space
 ${\bf {\Sigma}} = {\bf H}^2/{\Gamma}$ becomes a compact Riemann surface
 of genus greater than one. The idea is to choose a polygon bounded by
 geodesics as the fundamental region for the discrete group ${\Gamma}$,
 whose generators exhange pairs of edges of the polygon. In order to 
prevent that conical singularities arise in the quotient space the sum
 of the angles of the polygon has to be equal to $2{\pi}$. The simplest
 candidate for a polygon --- a square --- is ruled out because in the 
hyperbolic plane the sum of its angles is less than $2{\pi}$. To do the
 trick one needs a polygon with $4g$ sides, where $g \geq 2$. The sum 
of the angles can then always be set equal to $2{\pi}$ by adjusting the
 size of the polygon, since the angles shrink as the area of the 
polygon increases. The regular surface that arises when the edges have
 been identified is then a compact Riemann surface of genus $g$. 

The simplest possible case is that of a regular octagon with opposing 
edges identified, as illustrated in figure~\ref{fig5}. An elementary calculation
 shows that the Euclidean coordinate distance $d$ between the origin 
and the symmetrically placed edges has to be

\begin{equation} d = \sqrt{\sqrt{2} - 1} \label{17} \end{equation}

\noindent (in coordinates where the Euclidean coordinate radius of the 
disk is unity). 

So what we intend to do is to define a two-parameter family of 
Poincar\'{e} disks which is
 such that every point in the allowed region --- defined by eq. (\ref{14}) 
 --- lies on a unique disk, 
and such that each disk is mapped into itself by the $SO(2,1)$ group of 
isomorphisms generated by $J_{XU}, J_{YU}$ and $J_{XY}$. Then we select 
a discrete subgroup ${\Gamma}$ of $SO(2,1)$ and use it to compactify 
all the disks at one 
stroke. We have to check that all the compactified disks are smooth 
manifolds, so that the resulting solution will have the topology ${\bf R}^2 
\otimes {\bf {\Sigma}}$, where ${\bf {\Sigma}}$ is a Riemann surface
 of genus higher than one.

This is easier than it sounds. First we rewrite the equation that 
defines anti-de Sitter space as 

\begin{equation} X^2 + Y^2 - U^2 - V^2 = - (1 + Z^2) \ . \end{equation}

\noindent We see that $Z$ parametrizes a family of three-dimensional
 anti-de Sitter spaces that foliate the four-dimensional space. A 
coordinate system that takes advantage of this situation is

\begin{equation} Z = \sinh{z} \hspace*{8mm} V = \sin{v}\cosh{z} 
\hspace*{8mm} X = R\hat{X} \hspace*{8mm} Y = R\hat{Y} \hspace*{8mm}
 U = R\hat{U} \end{equation}

\noindent where 

\begin{equation} R = R(v, z) = \cos{v}\cosh{z} \end{equation}

\noindent and $(\hat{X}, \hat{Y}, \hat{U})$ obey eq. (\ref{16}). This 
coordinate system covers all of the allowed region (where the Killing 
vector fields we will use for identification are spacelike). In this 
region the space-time metric then takes the form

\begin{equation} ds^2 = - \cosh^2{z} dv^2 + dz^2 + R^2 d{\sigma}^2 \
 , \end{equation}

\noindent where $d{\sigma}^2$ is the metric on the hyperbolic plane 
defined by eq. (\ref{16}). Thus every point in the allowed region lies 
on a unique disk with a radius of curvature $R$ that depends on $z$ and 
$v$. Our $SO(2,1)$ Killing vectors are

\begin{equation} J_{XU} = X\partial_U + U\partial_X = 
\hat{X}\partial_{\hat{U}} + \hat{U}\partial_{\hat{X}} \end{equation}

\noindent and so on. Therefore they lie in the disks and the intersections
 of any disk with the level surfaces of the 
Killing vector fields are geodesics on the disk. Indeed 
figure~\ref{fig5} applies to all the disks if it applies to one of them, 
since the radius of curvature does not affect the coordinate distance 
$d$. This is all we need to see that our construction works; in particular
 conditions such as the condition on $d$ given in eq. (\ref{17}) 
will be fullfilled on all the disks if it is fullfilled on one.

To visualize the solution consider figure~\ref{fig6}, which shows the 
Poincar\'{e} ball defined by $t = 0$. It lies entirely within the 
allowed region. The identification surfaces are seen as segments of
 spheres going through the ball. Figure~\ref{fig6} illustrates the simplest 
case, where the identification surfaces have been chosen so that 
their intersections with the hyperbolic planes that foliate the ball
 form regular octagons. The compactified planes will then be compact
 surfaces of genus two. If we --- mentally --- add the fourth dimension
 to the picture every such compact surface can be thought of as an 
initial data surface for a solution of Einstein's equations in 2+1
 dimensions, giving rise to a locally anti-de Sitter space-time which
 begins and ends in a singularity.
\ \\

\noindent\parbox{\textwidth}{\noindent\begin{fig}{fig5}
  \mbox{\setlength{\epsfxsize}{.6\textwidth} \epsfbox{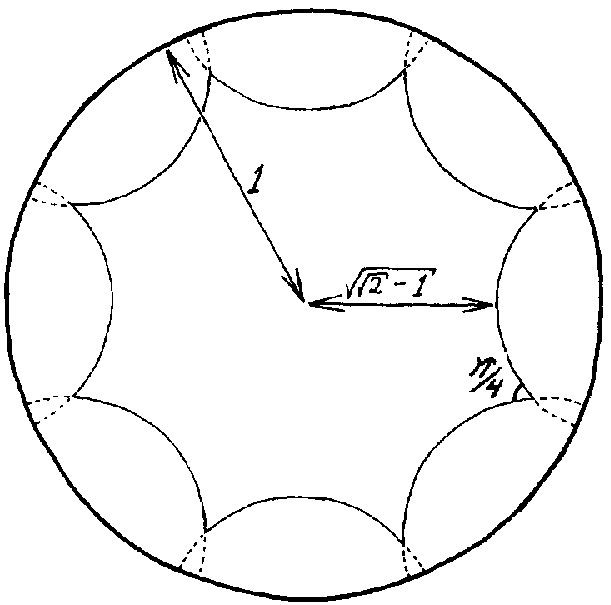} }\\
\ \\
  \figcap{.9\textwidth}{A regular octagon in the Poincar\'{e} disk such that its sum
    of angles equals $2\pi$. The edges are geodesics. 
    When opposing edges are identified we
    obtain a surface of genus two, with no singularities.}
\end{fig}
}\\

\noindent It remains to locate the event horizons. There are two asymptotic 
regions in the $\pm Z$ directions. A light ray from the origin has
 just enough time to reach infinity before the singularity at 
$t = {\pi}/2$ terminates the solution, and therefore the origin lies
 on an event horizon. It is obvious on symmetry grounds that the event
 horizon at $t = 0$ is precisely the Riemann surface that in figure~\ref{fig6}
 is depicted as the octagon that lies on the plane that goes through 
the middle of the ball. This event horizon then splits in two and 
moves outwards; the Penrose diagram for our solution is the same as 
the Penrose diagram for the BHTZ solution in figure~\ref{fig2}.

We have now completed the constructions that were announced in the 
abstract of our paper. Our first construction gave non-eternal black 
holes with toroidal event horizons and one asymptotic region, and the 
second gave eternal black holes with event horizons of genus higher 
than one and two asymptotic regions. (Actually a black hole with a 
toroidal event horizon can be constructed along the lines of the 
second construction as well, but it is an extremal black hole
 with one of the asymptotic regions ``replaced'' by a singularity.)
\ \\

\noindent\parbox{\textwidth}{\noindent\begin{fig}{fig6}
  \mbox{\setlength{\epsfxsize}{.67\textwidth} \epsfbox{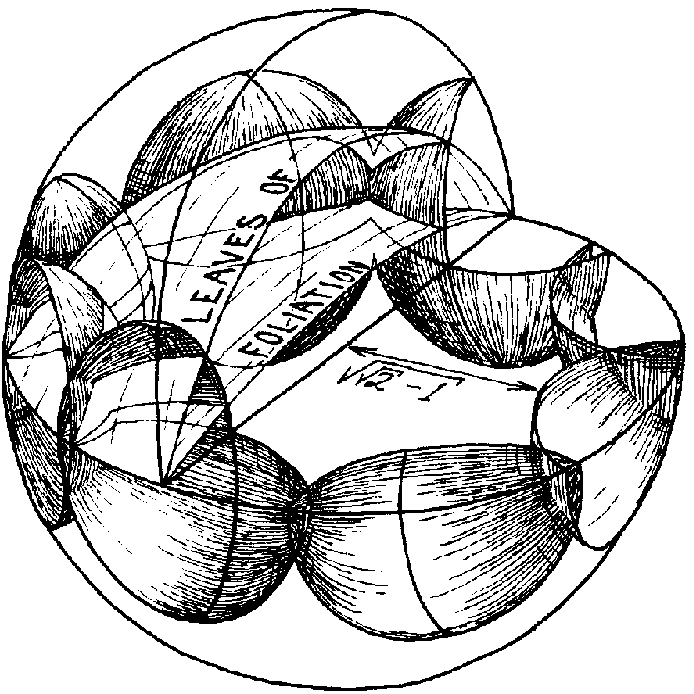} }\\
\ \\
\figcap{.9\textwidth}{The Poincar\'{e} ball at $t = 0$ (or equivalently $v = 0$).
    Now the identification surfaces are eight segments of spheres lying
    inside the ball and meeting its boundary at right angles. As
    before, opposing surfaces are identified. The ``leaves of
    foliation'' indicated in the figure are the surfaces of constant
    $Z$ at this time, all being the Poincar\'{e} disk of
    figure~\ref{fig5}. These leaves are segments of spheres, all
    containing the equator of the Poincar\'{e} ball.}
\end{fig}
}\\

\noindent We end with some comments about the possible significance of the black 
holes that we have made. First of all we see no way to obtain a black
 hole solution with the topology ${\bf R}^2 \otimes {\bf S}^2$ through 
identifications in anti-de Sitter space, and so we are unable to produce
 a constantly curved black hole with a physically sensible asymptotic
 behaviour. Therefore (and also because of the sign of the cosmological
 constant) we conclude that our constructions are of little direct 
relevance for physics. In particular they are of little relevance to the 
question of whether --- and if so, for how long --- event horizons of 
real black hole can be toroidal (see ref. \cite{Jacobson} for recent 
contributions). On the other hand our black holes appear to be close 
relatives to the one constructed by Lemos \cite{Lemos}, which also 
requires a space-time with the kind of asymptotic behaviour found in
 anti-de Sitter space. Whatever their direct relevance may be, we do 
believe that our constructions deserve attention as an amusing and 
perhaps an instructive footnote. Moreover, the occurence of event 
horizons in hyperbolic three-spaces is a subject of potential 
relevance for cosmology.\\
\ \\

\noindent {\bf Acknowledgements:}\\

\noindent Ingemar Bengtsson acknowledges Bo Sundborg for an 
instructive afternoon and the NFR for financial support.

\end{document}